# A Guide to Deal With Uncertainties in Software Project Management

Marcelo Marinho[1, 2] , Suzana Sampaio[2], Telma Lima[3] and Hermano de Moura[1]

[1]Informatics Center (CIn), Federal University of Pernambuco (UFPE), Recife, PE, Brazil
[2]Statistics and Informatics Department (DEINFO), Federal Rural University of Pernambuco (UFRPE), Recife, PE, Brazil
[3]Administration Department (DADM), Federal Rural University of Pernambuco (UFRPE), Recife, PE, Brazil

## ABSTRACT

*Various project management approaches do not consider the impact that uncertainties have on the project. The identified threats by uncertainty in a projec day-to-day are real and immediate and the expectations in a project are often high. The project manager faces a dilemma: decisions must be made in the present about future situations which are inherently uncertain. The use of uncertainty management in project can be a determining factor for the project success. This paper presents a systematic review about uncertainties management in software projects and a guide is proposed based on the review. It aims to present the best practices to manage uncertainties in software projects in a structured way including techniques and strategies to uncertainties containment.*

## 1. INTRODUCTION

The complexity and challenges involving software development, the use of techniques, practices and project management tools have become common in software engineering. Nowadays it is very common for companies to deal with software development or service provision as a project which needs to be planned, organized, conducted, monitored and controlled.

However, IT projects are notoriously disaster-prone, not necessarily because of technological failure but more often due to their inherent complexity. The Standish Group [1] were reporting that only 39% of projects, on average are delivered on time, within budget and with the agreed requirements (therefore those projects perceived as successful). 43% are delivered late, and/or over budget and/or under certain conditions and finally, 18% are cancelled on delivery and never used. In more than a decade, little seems to have changed.

Many projects with all the ingredients for success fail. It happens because executives, managers and project teams are not used to evaluating the uncertainties and complexities involved beforehand, and fail to adapt their management style to the situation [2].

Most projects face restrictions regarding time, costs and scope as well as certain criteria concerning to quality. Additionally, there is a high uncertainty level which can have both positive and negative effects on any project. The traditional approach to project management still emphasises assuring compliance with time, budget and scope constraints. Moreover, in project risk management literature, there is no common understanding as to what uncertainty is [3].

Uncertainty has no independent existence, it is not an object that can be identified and eliminated in the same way that a virus which invades a project can. The uncertainty arises naturally from complex situations, being only an inevitable factor of most projects. Uncertainty is simply an ambiguity expression and project indeterminacy in the same way that yellow is a colour attribute of daffodils, but is not a discrete or a separable part of the flower [4].

Aiming to better understand and explore the topic, a systematic review has been prepared. It is a planned and ideally repeatable way of synthesizing results from the existing body of scientific literature. It proceeds by discovering, evaluating and interpreting all available research relating to a particular question.

A systematic review process has three main phases: planning, conducting and reporting the review. The authors developed it with related studies from 1994 to 2013. Guided by research questions, the research aims to investigate: what the best practices to manage the uncertainties in software projects are; what the sources of uncertainty perceived by studies are; and what techniques or strategies are used for the recognition of the problem nature and containment of uncertainties in projects. To answer those core questions, the study dismembered the same three research questions that guided the work:

- **Research question 1:** How is it possible to reduce the uncertainty level in software projects?
- **Research question 2:** What techniques or strategies can help reduce the uncertainties in project management software?
- **Research question 3:** What are the sources of uncertainty perceived?

This research helps identifying the difficulties and the actions that may minimize the uncertainty effects in projects and how managers and teams may prepare themselves for the challenges in their project scenarios, based on the systematic review which was done. It has been created a uncertainty management guide for software projects to support project managers and teams in their day-by-day so it may reduce the uncertainty level in the project.

Besides the introductory section, this paper is structured as follows: Section 2 presents the systematic review process adopted for this study; Section 3 describes a data analysis extracted from the selected studies; in Section 4 the results for each research question are presented and summarized; Section 5 offers an overview of a guide for uncertainty management in software projects and Section 6 contains the conclusion.

## 2. SYSTEMATIC REVIEW PROCESS

Systematic literature reviews evaluate evidence in a systematic and transparent way. In a traditional literature review, the research strategy and results evaluation criteria are usually hidden from the reader, which means that the revision may perfectly be done in an unstructured way, ad hoc and evidence that do not support the researcher's preferred hypothesis might be ignored. However, in a systematic literature review the research strategy and the evaluation criteria are explicit and all relevant evidence are included in the evaluation [5],[6],[7].

This section describes the course of each step in the methodology used to carry out this systematic review study. We followed Kitchenham's methodological guideline for systematic reviews [8]. A systematic review protocol was written to describe the plan for the review. Details on the course of these steps are described in the following subsections.

### 2.1. Search environment

Before starting the researches, we decided to create a directory in the cloud. A free web store service was used by all researches to store all artifacts used; for example, electronic versions of publications, generated datasheets, partial reports and other documents. This enabled a total standardization and control of artifacts, so all researchers could access the artifacts as if they were in a local environment, thought they were remote.

Furthermore, we developed some datasheets to be used in all phases. The datasheets facilitated the organization of data in many aspects, for example a standard to enumerate publications searched, filters to extract objective information, the access of data of reasons by study and publications, and more. The datasheets also enabled the future access of data more precisely, where each phase could depend of other phase.

After this, we decided create responsibilities to researchers during the study, where the researchers should have responsibilities during to search and in each phase of search. The responsibilities were configuration management, artefacts development, analyzers and synthesizers. The responsibility that was aggregated to all researches was Analyzers.

## 2.2. Search strategy and search

A systematic review incorporates a search strategy for a research aiming to identify and retrieve even the slightest possibility of publications superset which meet the systematic review eligibility criteria. They are conditions to determine if primary studies are about the systematic review research questions. The search results are transformed into in a sequential publication list of the chosen engines. Each resource has a different community with differing interests, using different language and examining different issues. The engines provide different search syntaxes as well. Therefore, different resources might require different search strings.

After that, we conducted initial studies for all phases of the major study, that we called "pilot studies". These were performed to align a phase-to-phase understanding among researchers, all search engines mechanisms test and the adjust of some search terms. Only IEEE Explore search engine showed problems, which were solved with simply adjustments in the search terms for adapting to the search engine mechanism. The study only proceed when the two researchers agreed with the pilots results.

The resources used to searches are: IEEEXplore Digital Library (httt://ieeexplore.ieee.org/); ACM Digital Library (http://portal.acm.org); Elsevier ScienceDirect (www.sciencedirect.com); Springer Link (http://link.springer.com).

To search all results from sources, all researchers grouped to search publications. The sources (engines) were divided among all. Each researcher was responsible to find results in your engine and, finally catalogued. Then, when was performed the search, where was identified 3044 publications, according to results from engines mechanisms. The searches results were extracted in Bibtex files to merge in the datasheet developed to consolidate all results from all engines. After exclude duplicated results from datasheet, we found 2933 articles to start the first phase.

## 2.3. Paper selection

The idealized selection process has two parts: an initial selection of document research results that could plausibly satisfy the selection criteria, based on a reading of the title and summary of the articles, followed by a final selection of the list of initially selected works that meet the selection criteria, based on a reading of the introduction and conclusion of the work. To reduce potential bias, the selection process was conducted in pair, in which both researchers individually worked on the inclusion or exclusion of the work and then compared spreadsheets. Divergences were discussed and thus a consensus was reached. If there was not a consensus, a third researcher and should be consulted in case of doubt, the work would be inserted in the list.

In the pilot study performed before the first phase beginning, the first ten results in all engines were catalogued and all group read the titles and the abstracts and discussed about them to calibrate comprehension. Other pilot study was performed having more five publications done, because the researches were not ready to continue after the first pilot. After a reliability agreement, the first phase initiated. Each researcher read the publications` titles and abstracts to select or exclude the publication. together, they discussed about their results to gather them together according to a new datasheet agreement. Out of the initial selection of 2,933 papers, 111 articles were selected to second phase.

After the first phase and before second phase, a new pilot study was done. Then, we selected a single article to be read by researchers team aiming a consensus for both. In this phase, the introduction and the conclusion should be read. Similarly to the first phase, each researcher read the articles individually and later discuss its results together. After phase two selection, the researches eliminated 88 and selected 23 papers to be read for data extraction phase.

## 2.4. Study Quality Assessment

During the data extraction phase, the methodological quality of each publication was assessed. One researcher performed the quality assessment. Three factors were assessed as follows, and were each marked **yes** or **no**: Does the publication mention the possibility of selection, publication, or experimenter bias?;Does the publication mention possible threats to internal validity? and Does the publication mention possible threats to external validity?

The quality assessment was made solely on the basis of whether the publication explicitly mentioned these issues. We did not make judgements about whether the publication had a "good" treatment of these issues. The results of the study quality assessment were not used to limit the selection of publications.

## 2.5. Data Extraction

Before this stage, a new pilot was done to calibrate this design. We selected two relevant articles found by the authors (relevant for better quality in defined criteria) and we compared the extraction data performed so far with our data extraction. Thus, a pilot was carried out with an article found by us with one of the 23 selected works. In the data extraction phase, researchers must read the papers selected for extracting structured information according to the datasheet model.

We had selected 23 works, but during the extraction phase, the extractors identified 2 articles that showed no relevant citations or possible reasons to be extracted, thus, there were 21 articles. For each publication there were extracted information about the attributes defined in the datasheet.
From each study, there was extracted a list of shares, where each share described answer a research question. Or else, each simple sentence that answered one or more research question was considered a quota. We had a total of 147 quotas extracted from 21 studies. These shares were recorded on a datasheet.

## 2.6. Data Synthesis

The data extraction stage was over. The two researchers worked on the synthesis work to generate combinations of quotas with answers of the research questions.

There was a good level of inter-rater agreement, differences in opinion were discussed in a joint meeting, and it was easily resolved without the need of involving a third researcher arbitrating, as planned.

## 3. RESULTS

This section describes the analysis of the data extracted from our selected studies. As already mentioned in the methodology Section 2 of this work the systematic review process adopted had four main stages: Data Search, Data Selection, Data Extraction, Quality Assessment and Data Synthesis.

### 3.1. Data Search

In the Data Search phase the searches were conducted in four sources. The Figure 1 shows the results obtained on each stage at systematic review process. The survey was conducted for the period being between 1994 and 2013.

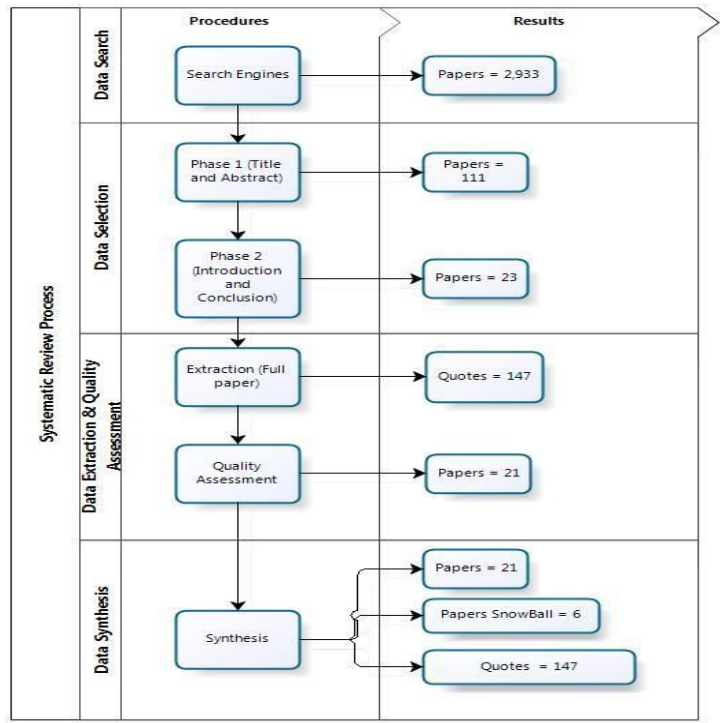

Figure 1. Results obtained on each stage at systematic review process.

Thus, were found the total of 3,044 papers in the period; at least 111 articles were identified and classified as duplicate articles. The search process was completed with a total of 2,933 articles ready for the next selection stage.

### 3.2. Data Selection

The Data Selection was divided in two phases: Phase 1: Title and Abstract analyses; and, Phase 2: Introduction and the Conclusion analyses.

In Phase 1, after checking the titles and abstracts, 111 articles were selected for the next phase. A total of 2,822 articles were eliminated. Among the criteria used, we may highlight: "0 - Not applicable" (this is a research non related with management or uncertainties), with 52%; "Outside the uncertainties in project management area", with 38% and "2 -It is about this risks in projects" with 10%.

In Phase 2, after reading the introduction and conclusion, just 23 papers were selected for the extraction phase. A total of 88 articles were eliminated. Among the criteria used we highlight: "1- Outside the uncertainties in project management area" with 59% , "2 - It is about this risks in projects" with 34%, followed by "0 - Not applicable", with 7%.

Table 1.  List of engines and its absolute contributions.

| Engine | Automatic | Selection 1 | Selection 2 | Extraction |
|--------|-----------|-------------|-------------|------------|
| ACM | 548 | 10 | 2 | 2 |
| IEEE | 722 | 63 | 15 | 13 |
| ScienceDirect | 569 | 11 | 4 | 4 |
| Springerlink | 1094 | 27 | 2 | 2 |
| Total | 2933 | 111 | 23 | 21 |

### 3.3. Data Extraction and Quality Assessment

In the Data Extraction and Quality Assessment each research read a full paper for the quotes extraction, at the same time they did the paper quality assessment. At the quality assessment phase three factors were assessed as follows, and were each researcher marked  YES or NO: i) Does the publication discuss the possibility of selection, experimenter, or publication bias?; ii) Does the publication discuss threats to internal validity?; and, iii) Does the publication discuss threats to external validity? The quality assessment was made only on the basis of whether the publication itself mentions with these issues – we did not assess whether the publications had a treatment of these issues. The Figure 5 shows a summary of all paper in this stage.

Out of 23 works selected in the previous stage, the researchers worked with 21 articles. Two were eliminated for not answering the research questions. Additional studies were identified by search technique snowball, or else, covering the studies references already found. This technique allowed us to identify high-quality works that were not found by the automatic search. 6 works that contributed to the discussion were added to be held in Section IV.  6 works out of 4  are books and 2 are journals that are not indexed by engines the selected research.

### 3.4. Data Synthesis

In the Synthesis phase 147 quotas were analyzed, in which 30 answer the first research question; 73 the second and 44 the last.

The geographical distribution of the uncertainties related to studies in project management  was as follows: The United States was ahead with 13 articles; England, with 3; Brazil, China and Singapore with 2; and Australia, Scotland, Philadelphia, Israel and Pakistan, with 1. Considering the studies evaluated in the extraction and adding the works found by snollball Figure 2 represents the distribution of works by country.

Figure 3 illustrates the distribution of studies identified by selection process throughout the years. We note that in the last 10 years it has been published 20 out of 27 of the papers included in the study. It demonstrates and confirms that research on uncertainties in project management have been growing since the last decade.

Figure 4 shows the distribution of documents found by search engines. Being 48% of the works found in IEEE Xplore.

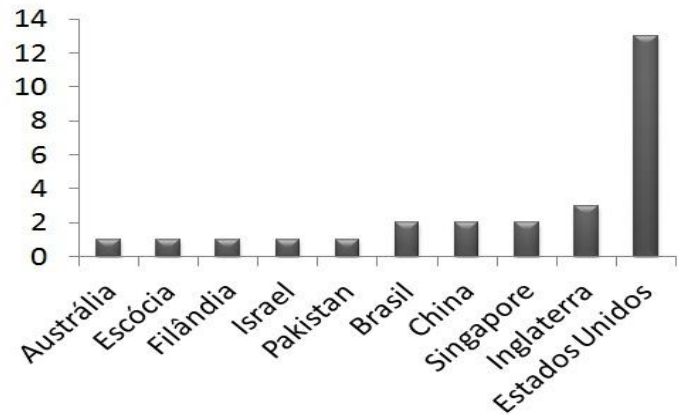

Figure 2. Distribution by country.

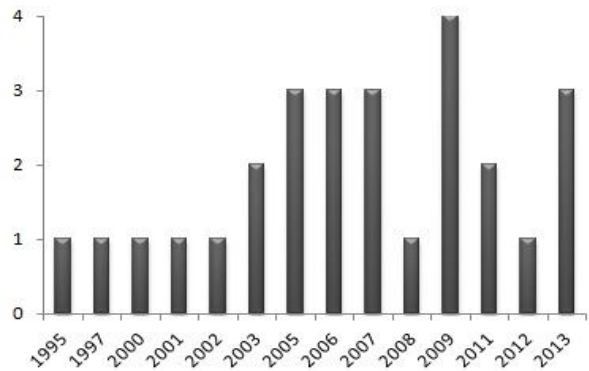

Figure 3. Distribution of studies identified by selection process throughout the years

Figure 5 shows the distribution of jobs by type of publication, showing that most of the studies, 44%, were published in journals, following a 41% of the annals of events (Conferences, Workshops and symposia) and 15% in books.

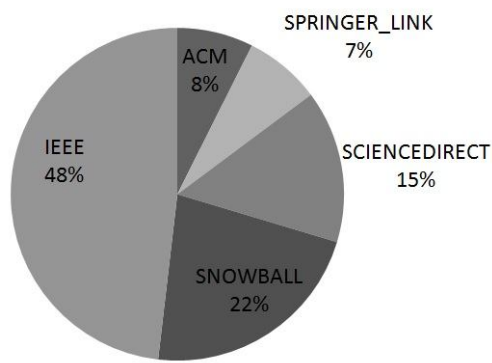

Figure 4. Distribution by engines

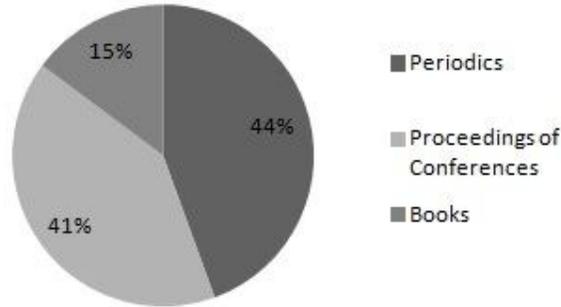

Figure 5. Distribution of jobs by type of publication

# 4. DISCUSSION

In this section, the results for each research question are presented. In Section 4.1 the evidences are presented about the possibility of reducing the uncertainties in software projects. In Section 4.2 evidences on how the techniques or strategies that favour uncertainties reduction in project management are presented. In Section 4.3 sources of uncertainties are perceived in the studies. In Section 4.4 the evidence that the uncertainties become larger and larger as the project is more innovative are presented. All evidence is properly referenced by 27 studies [9], [10], [11], [12], [13], [14], [15], [16], [17],[18], [19], [20], [21], [22], [2], [23], [3], [24], [25], [26], [27],[28], [29], [30], [31], [32], [33].

It is worth pointing out that some of the non- selected studies for extraction phase were considered so, because they confused risk with uncertainty. Note that although the risk and uncertainty are related, they are not the same thing. The uncertainty is the unknown, whereas risk is what can go wrong. Clearly, much of the risk of the project depends on the uncertainty, but there are other factors that contribute to the risk of the project, including deadlines, lack of resources and inadequate skills. Several of the articles report "uncertainties in the estimates" when in fact they are trying to estimate risks in the project process.

We made clear that the 27 studies agree on defining risk in technical terms as "state of knowledge in which each alternative leads to a result set, with the probability of occurrence of each result known by the decision maker". Uncertainty "the state of knowledge in which each alternative leads to a result set, with the probability of occurrence of each outcome is not known by the decision maker".

## 4.1. How is it possible to reduce the uncertainty level in software projects?

This question aimed to investigate the possibility of reducing the uncertainty level in software projects. From the 27 studies analyzed, 30 quotes were found and classified, there are 5 ways to manage uncertainties in projects identified by the research. They are: 9 approach adopting techniques and strategies to facilitate the uncertainties reduction; 8 address adapt management style to the projects type; 6 approach dealing with uncertainty when they happen; 5 approach understanding the uncertainty sources to better manage each type of project; 2 addresses identifying uncertainties in order to turn them into risks. Figure 6 presents the five practices summarized.

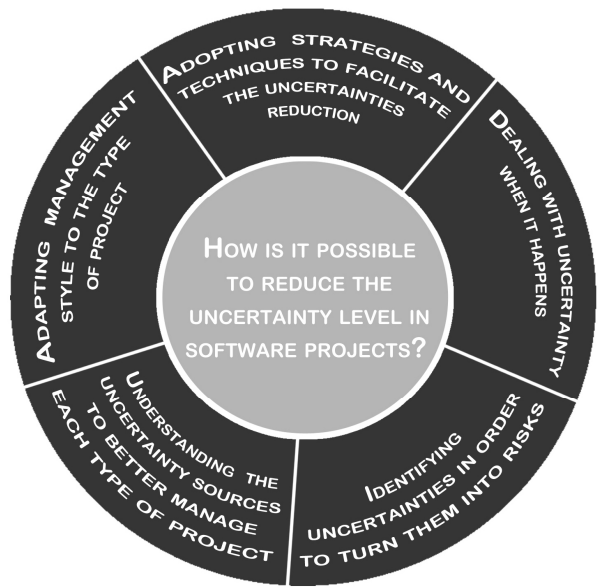

Figure 6. Ways to manage uncertainties in projects identified by the research.

### 4.1.1. Adopting techniques and strategies to facilitate the uncertainty reduction

It is tempting to want to eliminate all uncertainty, but the huge levels of necessary resources even to get close to that goal are, in all the most exceptional cases, unjustified. In fact, great efforts in the uncertainty sources eradication often divert attention from the real goals. Eradication is rarely the answer, it is more feasible to contain uncertainty within acceptable levels. This leads to another guiding principle for the uncertainty management: The objective is the uncertainty containment, not its elimination [13],[18], [22], [2], [23], [3], [29], [33].

Although there are no easy answers or quick solutions, the uncertainty can be ``tamed'', part of the answer lies in recognizing the nature of the problem and select the right technique (or strategy) to work. Like any good craftsman, the project manager must be in possession of a set of comprehensive tools for managing uncertainty and - equally important - a good knowledge of the capabilities and limitations of these tools [29]. For different types of problems, the manager and the team should have strategies, mindset and different paradigms [18].

### 4.1.2. Adapting management style to the type of projects

Many projects with all the ingredients of success still fail. The reason for this is that executives, project managers and the project team are not accustomed to assess and analyze uncertainties, and so fail to adapt their management style to the situation. Shenhar et al [34] affirm that "the same size does not fit all", meaning that every project is unique, and that one must understand how they differ and take the proper actions according to the particular needs of the organization and the project.

The authors describe the importance of evaluating and analyzing the uncertainties complexities of a project, and so adapt their management style to situation [31] . It is also addressed in some studies that projects fail because managers applied the wrong management style to the project [18]. It is important to point out that project managers cannot predict the future, but can perceive the uncertainty degree in their projects and choose an appropriate management style to manage them [10],[13].

### 4.1.3. Dealing with uncertainty when they happen

Some types of uncertainty cannot simply be solved through an analytical approach. It may be that a number of possibilities overlapping and combinations of random events contribute to deliver an unexpected result. Pharmaceutical companies have struggled with this problem. Despite extensive testing program, there is always the risk of an unlikely combination of environmental factors (other drugs usually administered by the patient) to react causing harmful side effects [21], [22], [2], [3].

Project managers can try to contain uncertainty at its source but will never have a hundred percent of success. Therefore, a project needs strength and should be able to rapidly detect and respond to unexpected events. For unexpected results a project manager must then decide how best to cope with uncertainty [29],[31].

### 4.1.4. Understand the sources of uncertainty to better manage each type of project

Uncertainty can arise from deficiencies in various sources, such as contextual information about the project, our understanding of the underlying processes, explanations of past events and the change speed (or time). But where these factors come into play within a project? What aspects of a project plan are particularly vulnerable to each uncertainty type? To answer these questions, first is interesting to see the elements that make up a typical project. Then, you need to examine what happens when the scale and complexity of the model increases. Thus, it is possible you can choose styles and strategies to manage the project properly [11], [13], [22], [15], [2], [23], [31].

### 4.1.5. Identify uncertainties in order to turn it into risk

Strategies can be used to contain the uncertainties. These strategies can help you learn more about the nature of uncertainty, for example, through the formulation of the problem that it represents or the modelling of future scenarios to prepare for them. Once an uncertainty is revealed, analytical techniques, such as risk management can be used in project management [11], [23].

## 4.2. What techniques or strategies can help reduce the uncertainties in project management software?

This question sought to identify techniques or strategies to support the software projects management that help reduce the uncertainties in innovative projects. Out of 73 quotas extracted for that matter, 18 were found techniques or strategies for managing projects focusing on reducing uncertainties. These techniques and strategies are described below with references of studies that support each of them.

### 4.2.1. Identifying the project type to adopt appropriate management

To reduce the probability of failure of a project it is important to characterize it properly, so knowing if there is a related uncertainty to their goals and solutions adopting a management model that fits the type project. Many projects with all the ingredients of success still fail. The reason for this is that executives, project managers and the project team are not accustomed to assess and analyze uncertainties, and so fail to adapt their management style to the situation [9], [10],[18], [19], [2], [23], [3], [29], [27], [28], [31], [33].

### 4.2.2. Managing the expectations of stakeholders so that they flexibly accept changes

Innovative projects can create high expectations for clients. You need to manage expectations, keeping them informed and aware of the project uncertainties, as well as creating a bond of trust between project members and clients [13], [18], [19], [21], [23], [3], [24], [27], [33].

### 4.2.3. Ability to formulate qualitative measures of success

Ability to formulate qualitative measures of success for projects is another tool that should be added to the arsenal of project management. Projects with low uncertainty can often be accessed through quantitative measures of success, such as time and cost, and tangible performance measures related to their tangible final deliverables. Projects with high levels of uncertainty require different forms of performance evaluation that recognize the validity of different perspectives and worldviews. This requires ability to develop assessment frameworks of sensitive performance that match the the project complexity, ie, It is necessary to understand what their contributions are to results in general, setting goals in advance for projects success, aligned with business goals of the organization [19], [2].

### 4.2.4. Identifying early warning signs to manage the uncertainties

Early identification of signs of a change can become a significant competitive advantage, because it can show an interruption in the current cycle, a break, beneficial or dangerous, for business. In the projects context, the early signs are of great importance ,especially in innovative projects by having several associated uncertainties. The idea is to identify the cause of problems, and the solutions for each signal [22], [3], [28], [33].

### 4.2.5. Sensemaking

The search for meaning is particularly important in a project-based environment. Once given way to a decision and its context, the actions (programs and projects), to be developed from the conception of meaning, become better understood and can be implemented in a more natural, efficient and effective[22], [3], [28], [33].

### 4.2.6. Management flexibility and ability to respond to changes

Complex and uncertain projects changes, requires greater flexibility and reflection, as a new way to generate knowledge, the managing way, the project manager and the performance of the team should change as the profile and the uncertainty evolve. Projects with many uncertainties must be open to creativity and experimentation. Thus, the flexibility and the ability to communicate the changes is fundamental [12], [13], [17], [20], [21], [2], [3], [31].

### 4.2.7. Managerial ability to perceive uncertainty and deal with them

The ability to take reasonable decisions to ensure there is necessary support to get everyone involved in the project; personal ability, such as intuition and trial to perceive uncertainties; ability to maintain a good relationship and build trust are favorable points for reduction and perception of uncertainty in a project [18], [12], [13], [21], [3], [26], [31].

### 4.2.8. Team willing to learn and develop new ideas in order to generate knowledge

Teams must go beyond mere crisis management and continuously observe the threats or opportunities. When new information arises, everyone should be willing to learn and then formulate new solutions [12], [13], [18], [3], [24], [31], [33].

### 4.2.9. The creation of flexible contracts

Creating flexible contracts for innovative projects help mitigating resistance to changes necessary for the project. Obviously, to have a flexible contract it is important to keep project stakeholders well informed [13],[33].

### 4.2.10. Building trust between team, management and customer

With uncertainty, a lot of time and effort should be to manage relationships with stakeholders and get them to accept unplanned changes. The relationship is characterized by trust between clients, managers and teams. Trust, once conquered, help alleviate the strategies change meetings during the project [13],[21],[28].

### 4.2.11. Verify information outside environment of the project

Actions relevant to organizations or groups within the organization (suppliers, competitors, consumers, government, shareholders, etc) can affect the product, as well as doubts about the likelihood or nature of changes in the environment general condition (socio-cultural trends, demographic changes) [13], [2], [28], [31], [32].

### 4.2.12. Understanding the sources of uncertainties

Project management can be conducted focused on resolving the project uncertainties, for doing so, it is necessary to understand where the uncertainties of projects can arise, ie, what are the possible sources of uncertainty? Understanding the sources we may be able to make the necessary changes as the project progresses [14], [2], [31], [32].

### 4.2.13. Project Managers must incorporate the investigation of uncertainties in projects

Ongoing investigation of uncertainties is important for members of the projects act in a proactive way and the organization to benefit strategically. The articles show that managerial knowledge aligned with the research uncertainties may contribute to transformation of uncertainties in risk [20], [2], [31], [32], [33].

### 4.2.14. Learning method

That includes experimentation and improvisation. The more we experience knowledge of a particular subject the more we reduce uncertainty; Improvisation itself can be a differentiator for innovative projects in the quest to deviate from uncertainties, seeking new goals [18],[28],[33].

### 4.2.15. Creativity techniques

Some articles suggest techniques such as: Brainstorming, feasibility study, market research to obtain knowledge [18],[28],[33].

### 4.2.16. Managers should facilitate communication within the organization

A propitious environment for communication can be a differentiator of organizations. Some articles suggest that innovative projects with small teams and located in the same environment have facility to pass the information received [13], [17], [18], [3].

### 4.2.17. Managers should facilitate self-organization and the team adaptability

They need to encourage diversity of thought and interaction, breaking organizational and hierarchical structures. The team needs to adapt to changes. They also need to interact feedback constantly with market and technology providers. The managerial focus should be on group dynamics to keep large project objectives in mind, instead of labor control [18], [28].

### 4.2.18. The Collaborative Work

The democratic management style is best appropriated, very tight control will lead an innovative project clutter and the project vision becomes an illusion. Collaborative working is essential in projects with many uncertainties [18], [2], [24].

### 4.2.19. Some techniques

Other techniques which help in reducing uncertainty were placed into just one article cited which were: quality agreement with the client, continuous integration, involvement of the specialist user in the project, short iterations [17]; Open mind to culture, stakeholder analysis; multidisciplinary team, multiple specialties together help in creating differentiated alternatives and phase uncertainties; external view of the problem reported by the team client [33].

## 4.3. What are the sources of uncertainty perceived?

This question sought to investigate what the sources of perceived uncertainty in software projects are. From the 27 studies analyzed, 44 quote was found. In the studies surveyed there is not a single label to the sources, it is common in some articles to speak from a particular source that is a single uncertainty, but classified differently, for example: market source [21], [17], [25], [22], [31], [32], external source [33] and novel source [11],[2] represent a single source of uncertainty. For doing so, we try to group the uncertainty and create a unique label for these sources and agree with Marinho et al [31] the classification of sources of uncertainty. Among those quotas found (they could cite more than one source, if they are in the same sentence), we find: 16 references to market uncertainty; 15 references to technological uncertainty; 14 for environment uncertainty and 9 for socio-human uncertainty. The following describes the sources of uncertainties grouped by research:

### 4.3.1. Market

Consumer market and different markets behave and think differently. Thus, project teams must know how their customers think, what their main problems are. Projects at various levels have their own unique elements that definitely stands out. If the market needs are already well known, the project will likely have little uncertainty. On the other hand, if they are not well understood it is necessary to guide the project to the desired goal. The market uncertainty comprehends client, suppliers, partners and current market situation [11],[13], [21], [17], [25], [22], [2], [29], [31], [32], [33].

### 4.3.2. Technological

It is the uncertainty source most cited in articles. The technological uncertainty depends on whether the project uses new technology or mature technology. The level of technological uncertainty of the project is not universal, but rather subjective, because it depends on technological know-how that exists or is accessible to the company. It is therefore a measure of the amount of existing new technology, compared to mature technology to use in the project. The technological uncertainty causes, among other things, an impact in the project, in the communication, in the freezing time of the plan and the number of planning cycles. It can also affect the technical expertise that the project manager and their team members need to have [10], [11], [13], [16], [17], [21], [22],[25], [31], [32].

### 4.3.3. Environment

The environment uncertainty can observe relevant actions to organizations or groups within the organization (suppliers, competitors, consumers, government, shareholders, team capacity, size of project lifecycle, resources, etc.) may affect the product, as well as doubts about the likelihood or nature of changes in the general condition in the environment (socio-cultural trends, demographic changes) [11], [16], [21], [29], [31], [32].

### 4.3.4. Socio-Human

Organizations have modern technological tools that meet the needs and structural deficiencies, but it is not enough to ensure individual and group knowledge acquisition due to cognitive factors intrinsically related to how people perceive, learn, remember and think about the information. Projects can be a unique way of organizational processes change, innovate and fit the reality of the competitive market, but project management can not be left to one person, it must become a question for everyone [26], [28], [30], [31].

## 5. GUIDE FOR THE MANAGEMENT THE UNCERTAINTIES

In their efforts to maintain their projects in good form, many project managers, implicitly, use similar ideas. However, they are not always as formal as the discipline demands. With this section it is hoped that organizations and managers formalize, explicitly, an approach directed towards the types of project which manage related uncertainties. This would be specifically for innovation projects.

While it is impossible to predict the nature of the problem in advance, project managers can use strategies that provoke a greater resistance in their projects. Throughout a project, we can establish a cycle in order to keep uncertainties managed, as shown in Figure 7.

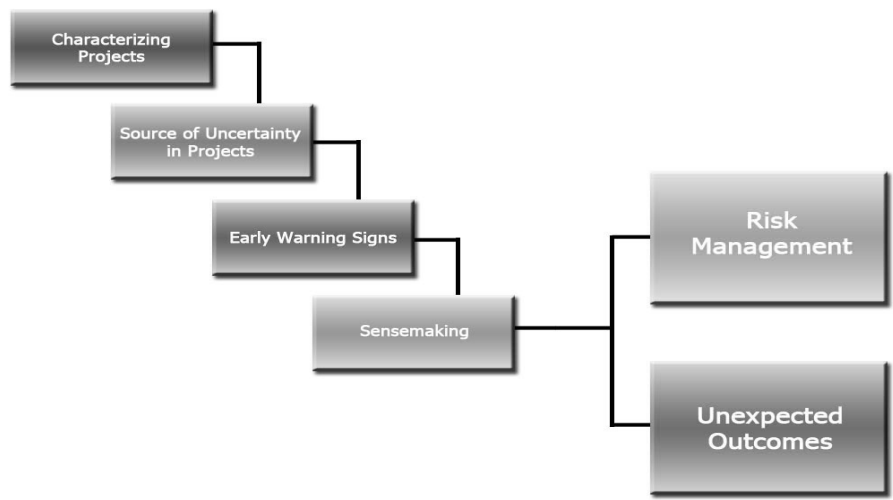

Figure 7. Guide for the Management of Projects.

The guide was elaborated based on the systematic review developed. The answers to the research questions presented and summarized in Section 4 were the base for the guide development. Note that the research question 4.1 presents ways the project managers are adopting to deal with the uncertainties. The guide took as reference the 5 ways (shown in figure 6 to establish a step-by-step guide. The adaptation of the management style is presented in the guide on Section 5.1; in Section 5.2 it is presented the identification phase of the uncertainty sources mentioned in the review as one of the uncertainty management ways; in Sections 5.3, 5.4, 5.5 the way how to transform uncertainties in risks is shown, it is presented an unexpected outcome phase in Section 5.6. Note that the adoption of techniques and strategies are mapped through the research question 4.2 and presented in the guide in Section 5.7.

In addition, the authors highlighted that in their studies, the techniques presented in Sections 5.3, 5.4 to transform uncertainties into risks were incorporated in Figure 7 for assuming that they are efficient practices to uncertainty management, however, not excluding any of the techniques or strategies presented in Section 4.2.

## 5.1. Characterizing Projects Phase

To reduce the probabilities of flaws in a project it is important to characterize it correctly, in this way coming to recognize whether or not an uncertainty exists in relation to aims and solutions, and adopting a model of management which is adequate for the type of project.

We can represent the characteristics of projects according to Figure 8. The first dimension is related to the objective of the project which could find itself with a level of certainty or uncertainty. The second dimension of the project refers to the solution, that is, whether there is certainty about the solution which should be elaborated. If we cross the dimensions, such as the ones presented in the Figure 8, we define a classification of which model can be adopted to manage the project. It is important to highlight that the barrier between what is clear or not clear is conceptual, meaning that it is not defined quantitatively. It is an intuitive classification to establish a better model for the management of the project.

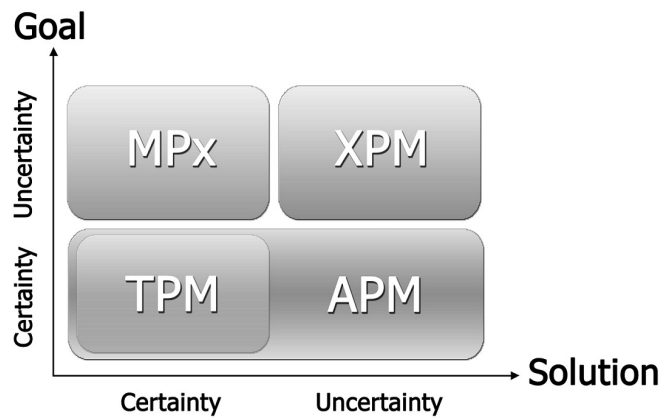

Figure 8. Characterizing Projects.

The projects can be classified as: **TPM** - Traditional Project Management; **APM** - Agile Project Management; **XPM** - Extreme Project Management; **mPX** - Emertxe Project Management;

The projects of the first quadrant of Figure 8can be executed by TPM or APM, being projects whose goals and solutions are clearly defined.

In the second quadrant there are projects with clear objectives and part of the unknown solution - immediately risk is related. For these projects there is the indicated utilization of APM, since the requisites are not defined in a clear way sufficiently to elaborate the complete planning of a Project, as happens with TPM.

In the third quadrant, even though it appears to make no sense at first sight, there are significant projects which consist of solutions looking for an objective.

In the fourth quadrant we have xPM projects. For this type of project there is great uncertainty in relation to the objectives and the solution. These are simultaneously apprehended and defined as part of the execution of projects. They are generally RD projects which run the risk of not having a conclusion. For these projects, the cycle of development can count on investigations and the construction of prototypes, all converging towards an objective which supports a solution.

## 5.2. Source of Uncertainty Phase

The uncertainty management starts with the understanding of the uncertainty sources. We may not always be aware of a specific uncertainty, but we can be alert to factors that may influence the success or failure of the project, it is important to understand the uncertainties sources discussed in Section 4.3.

The Figure 9 illustrates the four areas of uncertainty, which can be seen as a starting point for project managers to observe and identify uncertainties and thereby assist in the project's success. By accurately dimensional zing the categories and characteristics of uncertainty, large established corporations can better prepare project managers and senior leaders to anticipate and be sensitive to possible courses of evolution that projects may face.

## 5.3. Early Warning Signs Phase

The importance of weak signals was presented by Ansof [35], showing that the real world is full of information, which are often ambiguous, incomplete and inaccurate. Nevertheless, they may be transformed into significant advantages for companies. Many of them might be early signs of a break in the current cycle, either beneficial of dangerous for the business.

Several studies on early signs of project management deal with early symptoms trying to identify the necessary management actions for its management [36],[37]. The objective of this phase is to identify the signs of each project for uncertainties reduction.

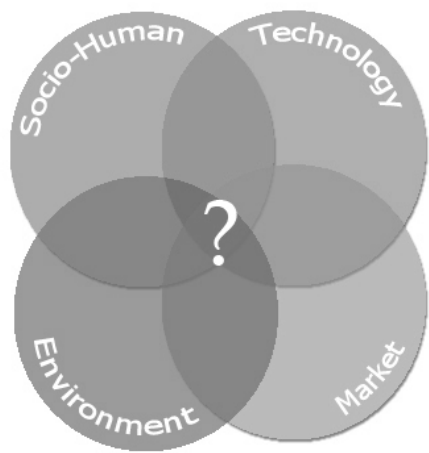

Figure 9. Source of Uncertainty in Projects.

### 5.4. Sensemaking Phase

Several studies on early signs of project management treat early symptoms trying to identify the necessary management actions. Practices such as Sensemaking are used in order to understand the signs in order to contribute to better project management [33]. The search for sense is particularly important in a project-based environment. Once certain sense is given to a decision and to its context, the actions (programs and projects), to be developed from the sense conception, become better understood and can be implemented in a more natural, efficient and effective.

Sensemaking is the process by which organizations and individuals work uncertainties, ambiguities, changes and problem situations generating inventions and new situations that result in actions that lead to problem solution and environmental stability. The most important thing is that there is sense in the identified signal or else, it is plausible to those involved [38],[28].

### 5.5. Risk Management Phase

If the signs are early detected and a sense for them is created, strategies can be used to contain the uncertainties. These strategies can help to learn more about the uncertainty nature, for example, through problem formulation by representing or modeling future scenarios and prepare for them. Using discovery techniques such as the construction of a knowledge map. Once uncertainty is revealed, analytical techniques such as risk management can be used in project management [39].

### 5.6. Unexpected Outcomes Phase

Project managers may try to contain the uncertainty in its source but will never be a hundred percent successful. Therefore, a project needs strength and should be able to detect and respond quickly to unexpected events. For unexpected results a project manager must then decide how best to cope with uncertainty:

- **Suppress**: It consists of strategies to reduce the uncertainty impacts, allowing the project to return gradually to the original plan;
- **Adapt**: A certain uncertainty level is accepted, however, one must be prepared to act quickly and limit the major impacts of any unexpected event;
- **Detour**: If possible, we should deviate from all uncertainty areas; unfortunately, deviate from them is not always possible. Some are unavoidable or too costly; one should be careful not to exchange an uncertainty for another; we can only deviate from what we know;
- **Reorient**: A more drastic deviation should be used only as a last resort; We must look for a different set of objectives for the project; used in cases when uncertainty drives the design to total failure.

### 5.7. Strategies and techniques

The strategies and techniques presented in Section 4.2 are presents in Figure 10, the idea is that the project manager may adopt the best practices to manage.

## 6. CONCLUSIONS

As the organizational environment becomes increasingly focused on projects, it is time to unleash the power and energy embedded in projects. The disturbingly poor project success rate makes it imperative for the organization to pay more attention to their project activity, their potential, and the competitive advantage they can bring. Unfortunately, even though the project management subject has changed a lot, it still remains focused on the parameters of time, budget and scope. It tends to miss the key point that projects always happen for commercial reasons.

The results of this research show that the number of papers on uncertainties related to project management has been growing since the last decade. There was also an increasing awareness of project management challenges. During the systematic review process some information was identified: the analyzed studies agree that uncertainties cannot be extinct from projects, but they may be "tamed". Our research has identified five good practices to manage the uncertainties in software projects; it was also investigated what the uncertainty sources are in the studies, being summarized into four; there were identified and catalogued eighteen techniques and/or strategies for the recognition of the problem nature and uncertainties containment in projects.

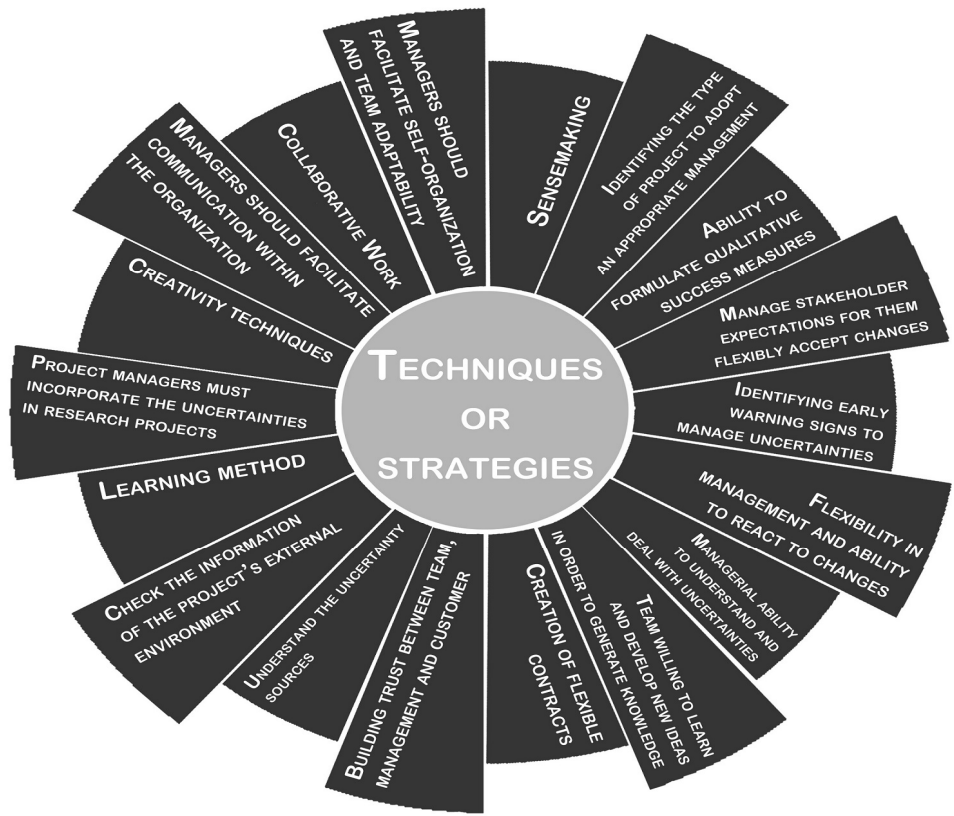

Figure 10. Strategies and techniques

In addition, it was presented a guide to manage the uncertainties in software projects. Practices and strategies in the guide were based on the research questions taken from the systematic review. The guide contains six phases which are: characterizing projects, uncertainty sources, early warning signs, sensemaking, risk management and unexpected outcomes.

One of the strengths of the proposed approach in this paper is that it is based on existing evidence in the industrial and scientific community, as these evidences were found in quality papers: journals, conferences and books. However, the guide preparation was based on evidence found in the systematic review and it is limited by the fact that it was not practically applied. As further work, the practical use of the guide through conducting case studies in industrial environments is necessary, not only to validate the guide, but also to increase the evidence number on the uncertainty management in projects. Te authors believe that the industrial application of the guide, can contribute to the improvement of the uncertainties reduction in projects, thus, contributing to project success in organizations.

## ACKNOWLEDGEMENTS

Authors would like to thank CAPES for supporting the development of this work as well as the reviewers.